\documentclass[preprint,showpacs,preprintnumbers,showkeys,amsmath,amssymb]{revtex4}

\usepackage{graphicx}
\usepackage{dcolumn}
\usepackage{bm}

\begin{document}


\title{Anomalous propagations of electromagnetic waves in anisotropic media with a unique dispersion relation}

\author{Weixing Shu}\thanks{Corresponding author. $E$-$mail$ $address$: wxshuz@gmail.com.}
\author{Hailu Luo}
\author{Fei Li}
\author{Zhongzhou Ren}
\affiliation{ Department of Physics, Nanjing University, Nanjing
210008, China}
\date{\today}

\begin{abstract}
We investigate the reflection and refraction behaviors of
electromagnetic waves at the interface between an isotropic
material and the anisotropic medium with a unique dispersion
relation.  We show that the refraction angle of whether phase or
energy flow for $E$-polarized waves is opposite to that for
$H$-polarized waves, though the dispersion relations for $E$- and
$H$-polarized waves are the same in such anisotropic media. For a
certain polarized wave the refraction behaviors of wave vector and
energy flow are also significantly different. It is found that
waves exhibit different propagation behaviors in anisotropic media
with different sign combinations of the permittivity and
permeability tensors. Some interesting properties of propagation
are also found in the special anisotropic media, such as
amphoteric refraction, oblique total transmission for both $E$-
and $H$-polarized waves, and the inversion of critical angle. The
special anisotropic media can be realized by metamaterials and
lead to potential applications, such as to fabricate polarization
splitters with higher efficiencies than conventional counterparts.
\end{abstract}

\maketitle

\section{Introduction}\label{Introduction}
Over thirty years ago, Veselago pioneered the concept of
left-handed material (LHM) with simultaneously negative
permittivity and negative permeability \cite{Veselago1968}. In
this kind of material, ${\bf k}$, ${\bf E}$ and ${\bf H}$ form a
left-handed set of vectors, and the material is thus called
left-handed material. It leads to many exotic electromagnetic
properties among which the most well-known is negative refraction
and can be used to fabricate a perfect lens \cite{Pendry2000}.
Since the negative refraction was experimentally observed in a
structured metamaterial formed by split ring resonators and copper
strips \cite{Shelby2001}, the LHM has attracted much attention and
interest. By now, more experiments \cite{Parazzoli2003,Houck2003}
and numerical simulations
\cite{Kong2002,Smith2002,Markos2002,Foteinopoulou2003} have added
new proofs to the negative refraction. Moreover, it has been found
that photonic crystals
\cite{Notomi2000,Luo2002a,Luo2002b,Luo2002c,Moussa}, optical
crystals \cite{Zhang2003} and anisotropic media
\cite{Lindell2001,Hu2002,Zhou2003,Luo2005} can exhibit negative
refraction except for the isotropic material, so the concept of
LHM should be extended.

Recently, an increasing amount of effort has been devoting to the
study on negative refraction in anisotropic media. Lindell
\textit{et al} firstly studied the negative refraction in
uniaxially anisotropic media in which the permittivity tensor
{\boldmath$\varepsilon$} and the permeability tensor
{\boldmath$\mu$} are not necessarily negative \cite{Lindell2001}.
In Refs.~\cite{Hu2002,Zhou2003,Shen2005}, the general behavior of
wave propagations in uniaxial media is investigated in detail.
These researches are mainly concentrated on isotropic and uniaxial
media, but less work has been done on negative refraction in
generally anisotropic media. In addition, only one kind of
polarized waves, \textit{e.g.}, $E$-polarized waves, is considered
in the previous work on negative refraction in generally
anisotropic media \cite{Smith2004,Grzegorczyk2005a,Smith2003}.
Then, one enquires naturally: How on earth are $H$-polarized waves
refracted in anisotropic media? Whether $H$-polarized waves are
refracted anomalously or regularly when $E$-polarized waves
exhibit negative refraction?

In this paper, we investigate the propagation of electromagnetic
waves in the anisotropic media with a uniform dispersion relation
for any polarized waves. We demonstrate that the refraction
behaviors of both phase and energy flow for $E$-polarized waves
are different from those for $H$-polarized waves, though the
dispersion relations for $E$- and $H$-polarized waves are the
same. Moreover, for a certain polarized wave the refraction
behaviors of wave vector and energy flow are significantly
different. We also find other interesting characteristics of wave
propagation, including oblique total transmission for both $E$-
and $H$-polarized waves and the inversion of critical angle.  The
special anisotropic media can be realized by metamaterials and can
lead to potential applications, such as to fabricate polarization
splitters with a higher efficiency than conventional counterparts.
Our results show that it is necessary to study the propagation of
phase and energy flow for both $E$- and $H$-polarized waves in
order to obtain a complete knowledge on characteristics of wave
propagation in anisotropic media.

This paper is organized in the following way. Section II gives the
unique dispersion of any polarized waves in the special
anisotropic media. We discuss the reflection and refraction at the
interface between an isotropic regular material and  the special
anisotropic medium in Sec.~III. Section IV shows in detail that
the reflection and refraction characteristics of waves are
different in the anisotropic media with different sign
combinations of $\boldsymbol{\varepsilon}$ and $\boldsymbol{\mu}$,
gives corresponding numerical results, and discuss how to realize
the media. A summary is given in Sec. V.

\section{Dispersion relations}\label{sec2}

In this section we present the dispersion relation of
electromagnetic wave propagation in the special anisotropic media.

For simplicity,  we assume the permittivity and permeability
tensors of the anisotropic media are simultaneously diagonal in
the principal coordinate system,
\begin{equation}
\boldsymbol{\varepsilon}=\left(
\begin{array}{ccc}
\varepsilon_x  &0 &0 \\
0 & \varepsilon_y &0\\
0 & 0 & \varepsilon_z
\end{array}
\right), ~~\boldsymbol{\mu}=\left(
\begin{array}{ccc}
\mu_x  &0 &0 \\
0 & \mu_y &0\\
0 &0 & \mu_z
\end{array}
\right).\label{matrix}
\end{equation}
Note that the anisotropic medium  can be realized by metamaterials
composed of periodic arrays of split-ring resonators and
conducting wires \cite{Smith2003,Smith2004a,Grzegorczyk2005b}, or
of periodic inductor-capacitor loaded transmission line circuits
\cite{Eleftheriades2002,Caloz2003,Caloz2004,Grbic2003,Iyer2006,Cui2005,Feng2005}.
In order to disclose the basic characteristics of the propagation
of waves, we do not consider losses here as in
Refs.~\cite{Lindell2001,Hu2002,Smith2004}, though no practical
material is ideally lossless. In fact, the metamaterial of low
losses has been realized \cite{Feng2005}. To simplify the
analyses, we assume the medium in a Cartesian coordinates $(O, x,
y, z)$ coincident with the principal coordinate system and
$\textbf{e}_x$, $\textbf{e}_y$, $\textbf{e}_z$ are unit vectors
along the axes.

Let us consider a plane wave of angular frequency $\omega$
propagating from an isotropic regular material into the special
anisotropic medium. We assume that the electric field is ${\bf
E}={\bf E_0}e^{i{\bf k\cdot r}-i\omega t}$ and that the magnetic
field is ${\bf H}={\bf H_0}e^{i{\bf k \cdot r}-i\omega t}$. The
dispersion relation of plane wave in the isotropic regular medium
is
\begin{equation}
k_{x}^2+ k_{y}^2+ k_{z}^2= \varepsilon_I
\mu_I\frac{\omega^2}{c^2}, \label{D1}
\end{equation}
where $k_x$, $k_y$, $k_z$ are the components of the ${\bf k}$
vector, and $\varepsilon_I$ and  $\mu_I$ are the permittivity and
the permeability. For the  anisotropic media, if the condition
\cite{Chen,Shen2005}
\begin{equation}
\frac{\varepsilon_x}{\mu_x}=\frac{\varepsilon_y
}{\mu_y}=\frac{\varepsilon_z }{\mu_z}=C, \label{e_i/u_i}
\end{equation}
is satisfied, the dispersion relations of any polarized plane
waves are the same
\begin{equation}
\frac{q_{x}^2}{\varepsilon_z \mu_y}+\frac{q_{y}^2}{\varepsilon_z
\mu_x}+\frac{q_{z}^2}{\varepsilon_y \mu_x}= \frac{\omega^2}{c^2},
\label{D2}
\end{equation}
where $C$ is a constant, and $q_j$ denotes the component of the
wave vector ${\bf q}$ in the $j$ direction $(j=x, y, z)$. When
$C>0$, the anisotropic medium is regarded quasiisotropic because
$E$- and $H$-polarized waves in it exhibits the same propagation
behaviors, just as in isotropic medium
\cite{Shen2005,Chen,Luo2006}. In the present paper we are
interested in the anisotropic medium with $C<0$.  We shall show
that $E$- and $H$-polarized waves exhibits different propagation
behaviors. That is why such anisotropic media can not be regarded
as quasiisotropic. The surface of wave vectors decided by the
dispersion relation with $C<0$ in wave-vector space is always a
single-sheet hyperboloid as shown in Fig. 1.
\begin{figure}
\includegraphics[width=8cm]{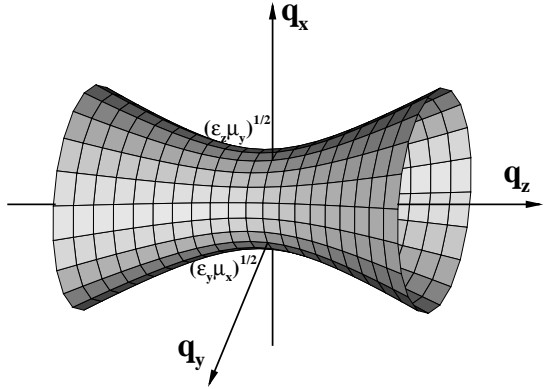}
\caption{\label{Fig0} The surface of wave vectors at a fixed
frequency determined by the dispersive relation Eq.~(\ref{D2})
with $C<0$ is always a single-sheet hyperboloid. Here, it has been
chosen that $\varepsilon_z \mu_y>0$, $\varepsilon_z \mu_x>0$,
$\varepsilon_y \mu_x<0$, and $\omega/c$ is the unit of coordinate
axes.}
\end{figure}

For simplicity we assume in the rest of the paper that the
incident, reflected and refracted waves are all in the $x$-$z$
plane, the boundary is at $z=0$ and the $z$ axis is directed into
the anisotropic medium. Then we can write the incident angle as
\begin{equation}\label{theta_I}
\theta_I =\tan^{-1}\left[\frac{k_x}{k_{z}}\right].
\end{equation}
The refractive angle of the transmitted wave vector or of the
phase is
\begin{equation}\label{refraction_angle}
\beta_P= \tan^{-1}\left[\frac{q_x}{q_{z}}\right].
\end{equation}
The refraction of phase is regarded regular if ${\bf k}_z\cdot
{\bf q}>0$ and anomalous if ${\bf k}_z\cdot {\bf q}<0$. In general
the occurrence of refraction requires that the $z$ component of
the refracted wave vector must be real. According to
Eqs.~(\ref{D1}) and (\ref{D2}) we obtain
\begin{equation}
{\varepsilon_y \mu_x}\left(\frac{k^2}{\varepsilon_I
\mu_I}-\frac{q_{x}^2}{\varepsilon_z \mu_y}
 \right)\geq0,\label{critical condition}
\end{equation}
where $k^2=k_x^2+k_z^2$. At the same time, we have
\begin{equation}
 q_z=\sigma\sqrt {{\varepsilon_y \mu_x}\left(\frac{\omega^2}{c^2}-\frac{q_{x}^2}{\varepsilon_z \mu_y}
 \right)},\label{qz}
\end{equation}
where $\sigma=+1$ or $\sigma=-1$. The choice of the sign should
ensure that the  power of electromagnetic waves propagates away
from the surface to the $+z$ direction.

\section{Reflection, refraction and Brewster angles}\label{sec3}
In this section we  discuss the reflection and refraction between
an isotropic regular material and the special anisotropic medium.
We  show that the refraction angles of both phase and energy flow
for $E$-polarized waves are opposite to the corresponding parts
for $H$-polarized, though the dispersion relations for $E$- and
$H$-polarized waves are the same in the special anisotropic
medium.

For E-polarized plane wave the electric fields can be expressed as
\begin{eqnarray}
&&{\bf E}_I=E_0{\bf e}_y e^{ik_x x+ik_z z-i\omega t},\\
&&{\bf E}_R=R_EE_0{\bf e}_y e^{ik_x x-ik_z z-i\omega t},\\
&&{\bf E}_T=T_EE_0{\bf e}_y e^{iq^{(E)}_x x+iq^{(E)}_z z-i\omega
t},
\end{eqnarray}
for incident, reflected and  refracted waves, respectively. Here
$R_E$ and $T_E$ are the reflection and transmission coefficients,
and the components of refracted wave vector $(q^{(E)}_x,
q^{(E)}_z)$ are decided by the dispersion relation Eq.~(\ref{D2}).
By boundary conditions we find that $q^{(E)}_x=k_x$ and  the
reflection and transmission coefficients are
\begin{eqnarray}\label{RE}
R_E=\frac{\mu_xk_z-\mu_Iq^{(E)}_z}{\mu_xk_z+\mu_Iq^{(E)}_z},
~~~T_E=\frac{2\mu_xk_z}{\mu_xk_z+\mu_Iq^{(E)}_z}.
\end{eqnarray}
The time-average Poynting vector is defined as ${\bf
S}=\frac{1}{2}Re({\bf E}\times {\bf H^*})$. For the refracted wave
we have
\begin{equation}
{\bf S}_T^{(E)}=Re \left[\frac{T_E^2E_0^2 k_x}{2 \omega \mu_z}{\bf
e}_x+\frac{T_E^2E_0^2q^{(E)}_z}{2\omega\mu_x}{\bf
e}_z\right].\label{SE_T}
\end{equation}
We should note that the sign of ${\bf q}^{(E)}\cdot{\bf
S}_T^{(E)}$ can not be used to demonstrate whether the refraction
of energy flow is regular or anomalous as in isotropic or uniaxial
anisotropic material \cite{Lindell2001,Hu2002}. Instead, we
consider the inner product of the tangential component of ${\bf
q}^{(E)}$ and ${\bf S}_T^{(E)}$
\begin{equation}
{\bf q}^{(E)}_x\cdot{\bf S}_{T}^{(E)}=Re \left[\frac{T_E^2E_0^2
k^2_x}{2 \omega \mu_z}\right]\label{k*SE_T}.
\end{equation}
The refraction of energy flow is regular if  ${\bf
q}^{(E)}_x\cdot{\bf S}_{T}^{(E)}>0$ and anomalous if ${\bf
q}^{(E)}_x\cdot{\bf S}_{T}^{(E)}<0$. Hence, the sign of $\mu_z$
indicates whether the refraction is regular or anomalous.

Following similar way, for $H$-polarized incident waves we obtain
the reflection and transmission coefficients
\begin{eqnarray}\label{RH}
R_H=\frac{\varepsilon_xk_z-\varepsilon_Iq^{(H)}_z}{\varepsilon_xk_z+\varepsilon_Iq^{(H)}_z},
~~~T_H=\frac{2\varepsilon_xk_z}{\varepsilon_xk_z+\varepsilon_Iq^{(H)}_z},
\end{eqnarray}
the Poynting vector is calculated to be
\begin{equation}
{\bf S}_T^{(H)}=Re\left[\frac{T_H^2H_0^2 k_x}{ 2
\omega\varepsilon_z}{\bf e}_x+\frac{T_H^2H_0^2
q^{(H)}_z}{2\omega\varepsilon_x}{\bf e}_z\right],\label{SH_T}
\end{equation}
and the inner product of the tangential component of wave vector
${\bf q}^{(H)}$ and ${\bf S}_T^{(H)}$ is
\begin{equation}
{\bf q}^{(H)}_x\cdot{\bf S}_{T}^{(H)}=Re\left[\frac{T_H^2H_0^2
k^2_x}{ 2 \omega\varepsilon_z}\right].\label{k*SH_T}
\end{equation}
Likewise, whether the refraction of energy flow for $H$-polarized
waves is regular is decided by the sign of $\varepsilon_z$.
Evidently, if $\mu_x\neq\mu_z$ or $\varepsilon_x\neq\varepsilon_z$
in Eqs.~(\ref{SE_T}) or (\ref{SH_T}), then a bending angle always
exists between ${\bf k}$ and ${\bf S}$, and therefore ${\bf k}$,
${\bf E}$ and ${\bf H}$ can not form a strictly left-handed
system. Further, ${\bf k}$ is not coincident with ${\bf S}$
judging by Eqs.~(\ref{k*SE_T}) and (\ref{k*SH_T}). This point can
be seen more clearly in the next section. The refraction angle of
the energy flow is defined to be $\beta_S=
\tan^{-1}\left[{S_x}/{S_{z}}\right]$. Using Eq.~(\ref{SE_T}) and
(\ref{SH_T}) we thus have
\begin{equation}\label{beta_S_EH}
\beta_S^{(E)}=\tan^{-1}\left[\frac{\mu_xk_x}{\mu_zq^{(E)}_z}\right],
~~~\beta_S^{(H)}=\tan^{-1}\left[\frac{\varepsilon_xk_x}{\varepsilon_zq^{(H)}_z}\right],
\end{equation}
for $E$- and $H$-polarized waves, respectively.

Let us  note that the refraction direction should obey two rules:
The tangential components of the incident, reflected, refracted
wave vectors on the interface are continuous; The Poynting vector
of the refracted wave points away from the interface. Embodied in
the problem we are concerned with, the two rules are: $q_x=k_x$
and $S_{Tz}>0$. Then, making use of Eqs.~(\ref{theta_I}) and
(\ref{critical condition}), we find there exists a critical value
\begin{equation}\label{critical angle}
\theta_C= \sin
^{-1}\left[\sqrt{\frac{{\varepsilon_z}\mu_{y}}{{\varepsilon_I}\mu_I}}\right]
\end{equation}
for the incident angle if
$0<{\varepsilon_z}\mu_{y}<{\varepsilon_I}\mu_I$. At the same time,
from Eqs.~(\ref{SE_T}) and (\ref{SH_T}), we conclude that
$q^{(E)}_z$ must have the same sign as $\mu_x$ for $E$-polarized
waves and $q^{(H)}_z$ must have the same sign as $\varepsilon_x$
for $H$-polarized waves. Therefore, the sign of $\mu_x$ suggests
whether the refraction of phase in Eq.~(\ref{refraction_angle}) is
regular or anomalous for $E$-polarized waves, and the sign of
$\varepsilon_x$ does the same for $H$-polarized waves. Considering
Eq.~(\ref{qz}) and $\varepsilon_x/\mu_x<0$, we come to the
conclusion
\begin{equation}\label{beta_E=beta_H}
\beta_P^{(E)}=-\beta_P^{(H)}.
\end{equation}
That is to say, the refraction angle  of the phase  for
$E$-polarized wave is always opposite to that for $H$-polarized
wave. Then, together with Eqs.~(\ref{e_i/u_i}), (\ref{beta_S_EH})
and (\ref{beta_E=beta_H}), we find
\begin{equation}
\beta_S^{(E)}= -\beta_S^{(H)},
\end{equation}
which means that the refraction angle of energy flow for
$E$-polarized wave is always opposite to that for $H$-polarized
wave.

We next explore the problem of oblique total transmission. The
reflectivity and the transmissivity are defined to be
\cite{Kong2000}
\begin{eqnarray}\label{rt}
r=-\frac{{\bf e}_z\cdot{\bf S}_R}{{\bf e}_z\cdot{\bf S}_I},
~~~t=\frac{{\bf e}_z\cdot{\bf S}_T}{{\bf e}_z\cdot{\bf S}_I},
\end{eqnarray}
respectively, where ${\bf S}_I$ and ${\bf S}_R$ are the respective
Poynting vectors of incident and reflected waves. Then we have for
$E$ and $H$-polarized waves
\begin{eqnarray}
r_E=|R_E|^2,~~~t_E=\frac{\mu_Iq^{(E)}_z}{\mu_xk_x}|T_E|^2;\label{rtE}\\
r_H=|R_H|^2,~~~t_H=\frac{\varepsilon_Iq^{(H)}_z}{\varepsilon_xk_x}|T_H|^2.\label{rtH}
\end{eqnarray}
It is easy to show that $r_E+t_E=1$ and $r_H+t_H=1$, which
indicates the power conservation on the boundary. When $r_E=0$ or
$r_H=0$, the incident angle is known as the Brewster angle
\cite{Grzegorczyk2005b}. Substituting Eq.~(\ref{RE}) into
Eq.~(\ref{rtE}), we find that when the condition
\begin{equation}\label{Brewster_E_condition}
0<\frac{\mu_z(\varepsilon_y \mu_I-\varepsilon_I
\mu_x)}{\varepsilon_I(\mu^2_I-\mu_x \mu_z)}<1
\end{equation}
is satisfied, the Brewster angle for $E$-polarized wave is
\begin{equation}\label{Brewster_E}
\theta_{B}^{(E)}=\sin ^{-1} \left[\sqrt{\frac{\mu_z(\varepsilon_y
\mu_I-\varepsilon_I \mu_x)}{\varepsilon_I(\mu^2_I-\mu_x
\mu_z)}}\right].
\end{equation}
Similarly, inserting  Eq.~(\ref{RH}) into Eq.~(\ref{rtH}) we can
see that under the condition
\begin{equation}\label{Brewster_H_condition}
0<\frac{\varepsilon_z(\mu_y \varepsilon_I-\mu_I
\varepsilon_x)}{\mu_I(\varepsilon^2_I-\varepsilon_x
\varepsilon_z)}<1,
\end{equation}
there exists a Brewster angle for $H$-polarized wave
\begin{equation}\label{Brewster_H}
\theta_{B}^{(H)}=\sin ^{-1} \left[\sqrt{\frac{\varepsilon_z(\mu_y
\varepsilon_I-\mu_I
\varepsilon_x)}{\mu_I(\varepsilon^2_I-\varepsilon_x
\varepsilon_z)}}\right].
\end{equation}
At the Brewster angle, the reflectivity is zero and the oblique
total transmission occurs. It is worthy of mentioning that such
oblique total transmissions for $E$- and $H$-polarized waves can
not occur simultaneously in a conventional isotropic material.
Therefore the oblique total transmission is due to the anisotropy
of the material  \cite{Zhou2003}.

\section{detailed analyses and numerical results}
In the following we shall apply the conclusions in the above
section to exploring in detail the reflection and refraction
characteristics at the boundary between an isotropic material and
the special anisotropic medium. Aggregately there are six kinds of
sign combinations for $\boldsymbol{\varepsilon}$ and
$\boldsymbol{\mu}$ of the anisotropic medium. We show that with
different sign combinations of $\boldsymbol{\varepsilon}$ and
$\boldsymbol{\mu}$, the behavior of propagation is different.
According to the curve form of refracted wave vector at a fixed
frequency (isofrequency curve) in the wave vector plane
\cite{Smith2004} we discuss in three cases.

\subsection{the isofrequency curve is a hyperbola with foci on the $k_x$ axis}
When the signs of the permittivity and permeability tensors are
chosen as $\boldsymbol{\varepsilon}=(-, -, +)$ and
$\boldsymbol{\mu}=(+, +, -)$, the isofrequency curve of the
refracted wave vector is a hyperbola with the foci in the $k_x$
axis, as shown in Fig. 2 (a).
\begin{figure}
\includegraphics[width=8cm]{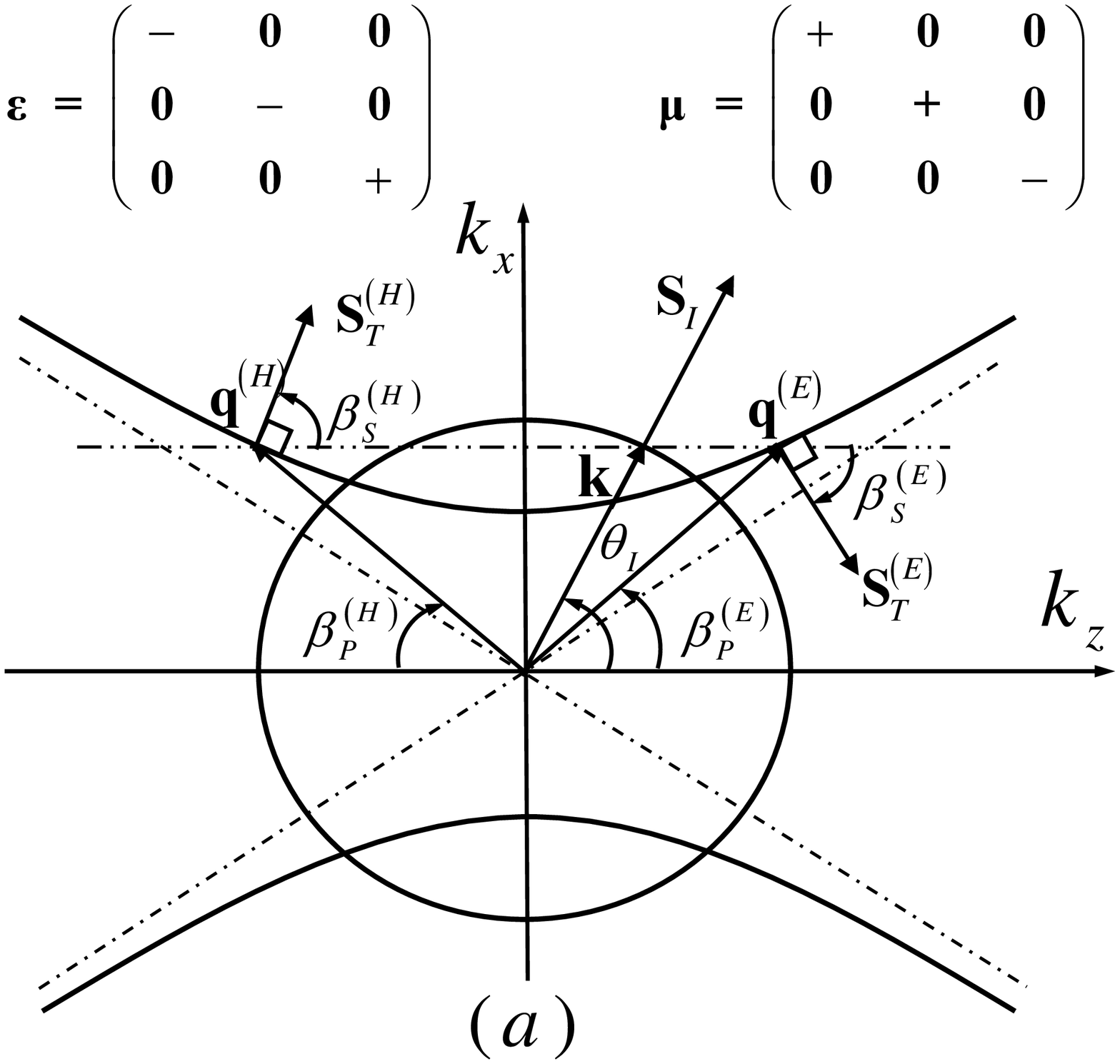}
\includegraphics[width=8cm]{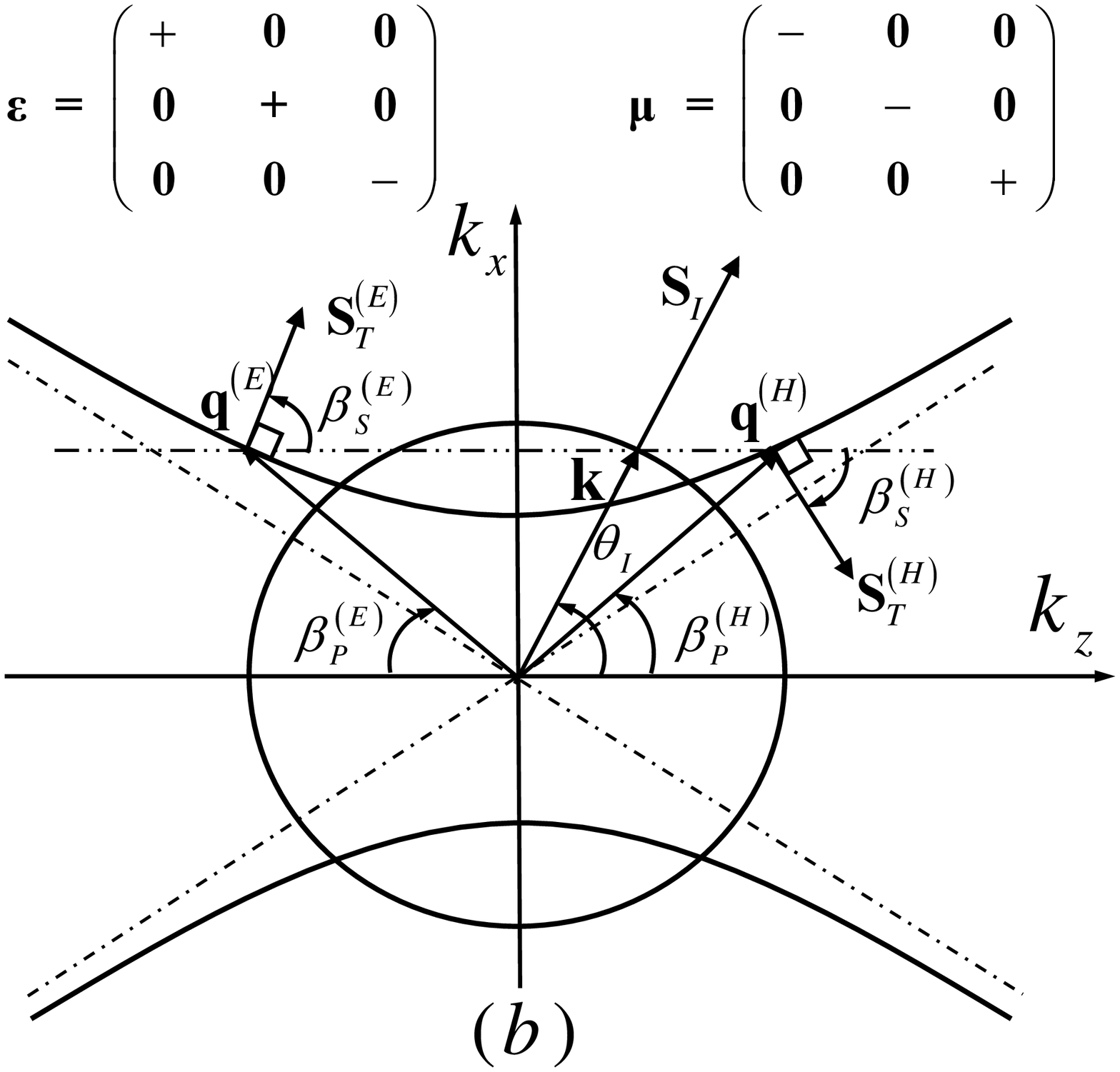}
\caption{\label{Fig1}The isofrequency curves to illustrate the
refraction. The circle and the hyperbola represent the surfaces of
wave vectors in an isotropic regular medium and the anisotropic
media, respectively. In (a),  $\beta_P^{(E)}=-\beta_P^{(H)}>0,
~\beta_S^{(E)}=-\beta_S^{(H)}<0$, while
$\beta_P^{(H)}=-\beta_P^{(E)}>0, ~\beta_S^{(H)}=-\beta_S^{(E)}<0$
in (b).}
\end{figure}
Note that this case corresponds to the \textit{anti-cutoff}
indefinite medium in Ref.~\cite{Smith2003}. Applying the
conclusions in the previous section to this case, we have
\begin{eqnarray}
&&\mu_x>0,~q^{(E)}_z>0;~\mu_z<0,~{\bf q}^{(E)}_x\cdot{\bf S}_{T}^{(E)}<0.\\
&&\varepsilon_x<0,~q^{(H)}_z<0;~\varepsilon_z>0,~{\bf
q}^{(H)}_x\cdot{\bf S}_{T}^{(H)}>0.
\end{eqnarray}
For the $E$-polarized wave the wave vector is refracted regularly,
but the Poynting vector is refracted anomalously. For the
$H$-polarized wave the situation is just opposite. More evidently,
\begin{equation}
\beta_P^{(E)}=-\beta_P^{(H)}>0, ~\beta_S^{(E)}=-\beta_S^{(H)}<0.
\end{equation}
If $0<{\varepsilon_z}\mu_{y}<{\varepsilon_I}\mu_I$, the refraction
can occur only if the incident angle is in the branch
\begin{equation}
\theta_C<|\theta_I|<\pi/2,
\end{equation}
where  $\theta_C$ is the critical angle defined by
Eq.~(\ref{critical angle}), or else $0<|\theta_I|<\pi/2$. That is
to say, the incident angle can be larger than the critical angle,
different from the regular situation where the incident angle is
smaller than the critical angle. At the same time, using
Eqs.~(\ref{critical angle}), (\ref{Brewster_E}) and
(\ref{Brewster_H}) one can easily show that the Brewster angle is
larger than the critical angle
\begin{equation}
\theta_{B}^{(E)}>\theta_C,~ \theta_{B}^{(H)}>\theta_C.
\end{equation}
This phenomenon is called the inversion of critical angle
\cite{Zhou2003,Grzegorczyk2005b}. From Eqs.~(\ref{rtE}) and
(\ref{rtH}) one can show that a bit shift of $\theta_I$ from
$\theta_C$ can lead to a big change in reflectivity, as can be
seen in Fig.~3.

If the signs of the permittivity and permeability are chosen as
$\boldsymbol{\varepsilon}=(+, +, -)$ and $\boldsymbol{\mu}=(-, -,
+)$, the iso-frequency curve of the refracted wave vector is shown
in Fig.~2(b). We can see that the refraction behaviors of $E$- and
$H$-polarized waves are just opposite to the counterparts in
Fig.~2(a), though the dispersion relation is invariant. This is
due to the sign inversions of $\boldsymbol{\varepsilon}$ and
$\boldsymbol{\mu}$ in Fig.~2(b) compared with those in Fig.~2(a)

For the purpose of illustration, a numerical example is given in
Fig.~3.
\begin{figure}
\includegraphics[width=8cm]{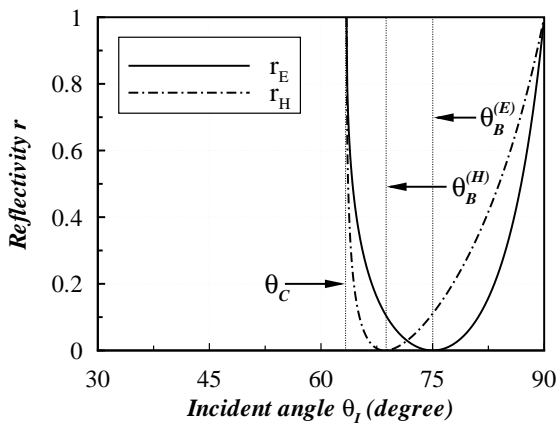}
\caption{\label{Fig1} Diagram of reflectivity as a function of
incident angle. For the isotropic material $\varepsilon_I=1$ and
$\mu_I=1$, while $\boldsymbol{\varepsilon}=(-0.5, -0.4, 1)$ and
$\boldsymbol{\mu}=(1, 0.8, -2)$ for the anisotropic medium. There
exist Brewster angles for both $E$- and $H$-polarized waves,
\textit{i.e.}, $\theta_{B}^{(E)}$ and $\theta_{B}^{(H)}$. The
$\theta_C$ is the critical angle.}
\end{figure}
We find that there exists a Brewster angle for both $E$- and
$H$-polarized waves and that the Brewster angles are larger than
the critical angle. This phenomenon is different from the
situation in conventional isotropic nonmagnetic materials where
the Brewster angle only exists for $H$-polarized waves. In a
regular nonmagnetic material, the physical mechanism of Brewster
angle is that the component of energies radiated by electric
dipoles under the effect of transmitted electric fields is zero
along the direction perpendicular to the reflected wave
\cite{Kong2000}. However, the appearance of Brewster angles here
relies on parameters of $\boldsymbol{\varepsilon}$ and
$\boldsymbol{\mu}$. First, let us consider an example: For a light
incident from a regular magnetic material with $\varepsilon=1$ and
$\mu=2$ into another regular material with $\varepsilon'=1$ and
$\mu'=1$, there exists a Brewster angle for $E$-polarized waves.
Evidently, the cause is not the response of electric dipoles, but
the response of magnetic dipoles \cite{Smith2004}. Therefore, the
existence of Brewster angles for $E$- and $H$-polarized waves here
is due to the compound operations of the electric and magnetic
responses in the anisotropic media.

\subsection{the isofrequency curve is an ellipse}
If the signs of the permittivity and permeability tensors have the
form of $\boldsymbol{\varepsilon}=(-, +, -)$ and
$\boldsymbol{\mu}=(+, -, +)$, the isofrequency curve of the
refracted wave vector is an ellipse as shown in Fig. 4 (a).
\begin{figure}
\includegraphics[width=8cm]{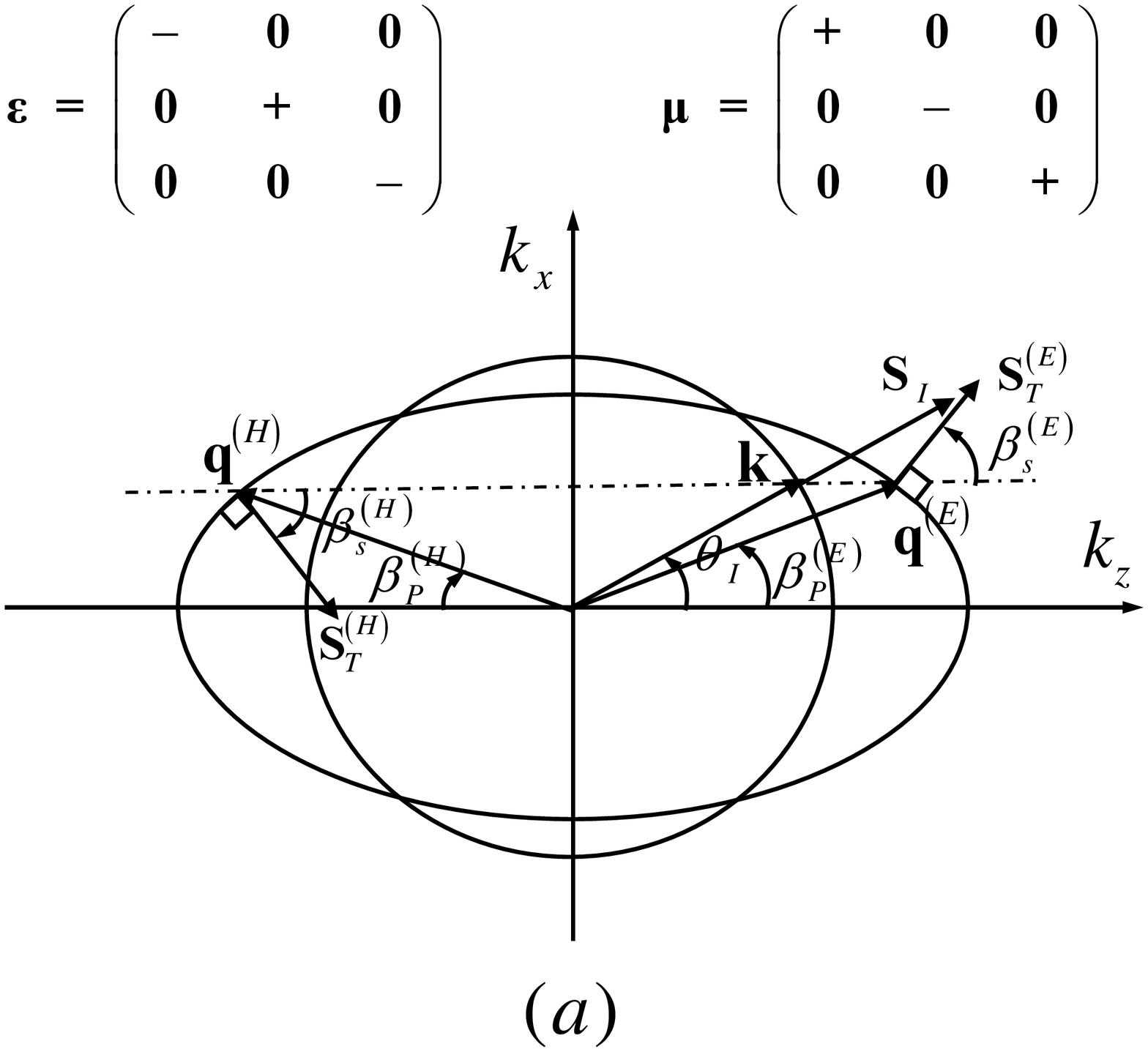}
\includegraphics[width=8cm]{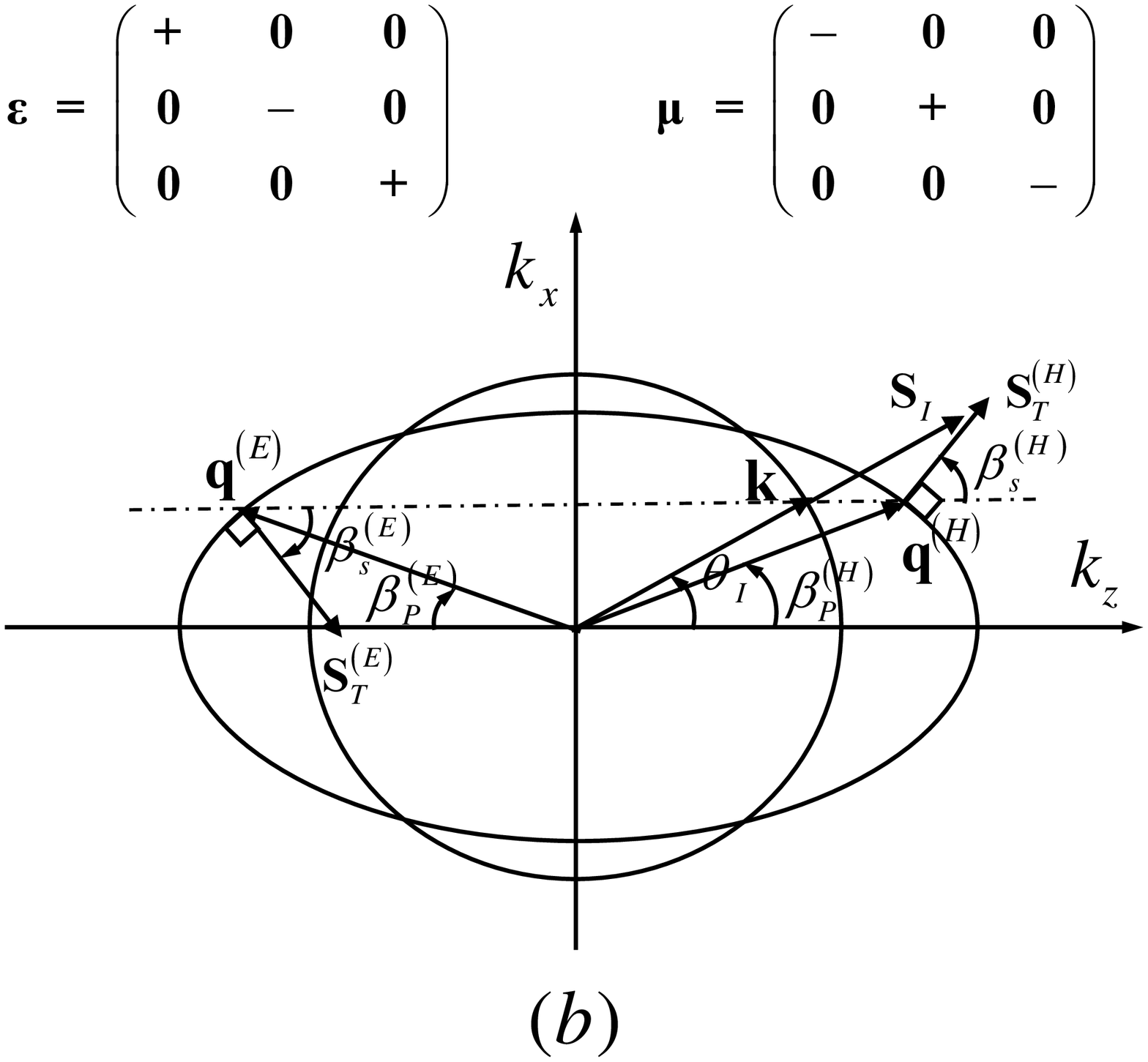}
\caption{\label{Fig3} The isofrequency curves to illustrate the
refraction. The circle and the ellipse represent the surfaces of
wave vectors in an isotropic regular material and the anisotropic
medium. In (a), $\beta_P^{(E)}=-\beta_P^{(H)}>0,
~\beta_S^{(E)}=-\beta_S^{(H)}>0$. The refraction situation is
reversed in (b), \textit{i.e.}, $\beta_P^{(H)}=-\beta_P^{(E)}>0,
~\beta_S^{(H)}=-\beta_S^{(E)}>0$.}
\end{figure}
This case corresponds to the \textit{cutoff} indefinite medium in
Ref.~\cite{Smith2003}. For this case we find
\begin{eqnarray}
&&\mu_x>0,~q^{(E)}_z>0;~\mu_z>0,~{\bf q}^{(E)}_x\cdot{\bf S}_{T}^{(E)}>0.\\
&&\varepsilon_x<0,~q^{(H)}_z<0;~\varepsilon_z<0,~{\bf
q}^{(H)}_x\cdot{\bf S}_{T}^{(H)}<0.
\end{eqnarray}
The wave vector and the Poynting vector are both refracted
regularly for the $E$-polarized wave, while anomalously for the
$H$-polarized wave. By virtue of the analyses in the previous
section, we obtain
\begin{equation}
\beta_P^{(E)}=-\beta_P^{(H)}>0, ~\beta_S^{(E)}=-\beta_S^{(H)}>0.
\end{equation}
From the figure we  can also see that, when
$0<{\varepsilon_z}\mu_{y}<{\varepsilon_I}\mu_I$, the refraction
can occur only if the incident angle is in the branch
\begin{equation}
-\theta_C<\theta_I<\theta_C,
\end{equation}
where  $\theta_C$ is the critical angle, or else
$-\pi/2<\theta_I<\pi/2$. In the case of
$\boldsymbol{\varepsilon}=(+, -, +)$ and $\boldsymbol{\mu}=(-, +,
-)$ the isofrequency curve of the refracted wave vector is shown
in Fig. 4 (b). We can see that the refraction behaviors of the
$E$- and $H$-polarized waves are just the results after exchanging
the $E$- and $H$-polarized waves in Fig. 4 (a). This result
manifests again that $E$- and $H$-polarized waves do not
necessarily exhibit the same propagation, even if their dispersion
relations are the same. As an example, we examine the case of
$\boldsymbol{\varepsilon}=(-0.5, 0.4, -1)$ and
$\boldsymbol{\mu}=(1, -0.8, 2)$, with the results illustrated in
Fig.~5. We find that there exists a Brewster  angle for
$H$-polarized waves. Actually, we can choose appropriate material
parameters of $\boldsymbol{\varepsilon}$ and $\boldsymbol{\mu}$
for $E$-polarized waves to exhibit a Brewster angle.
\begin{figure}
\includegraphics[width=8cm]{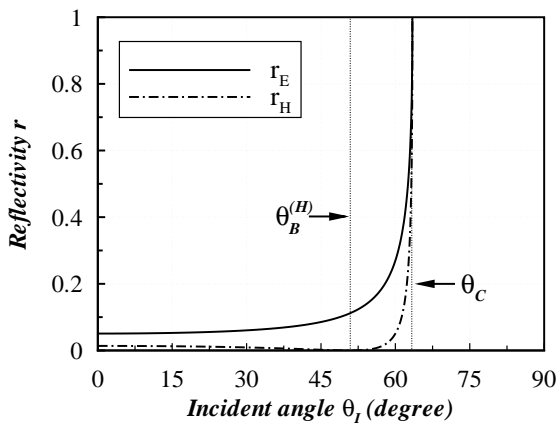}
\caption{\label{Fig4} Diagram of reflectivity as a function of
incident angle. Here $\varepsilon_I=1$, $\mu_I=1$,
$\boldsymbol{\varepsilon}=(-0.5, 0.4, -1)$ and
$\boldsymbol{\mu}=(1, -0.8, 2)$. The $\theta_{B}^{(H)}$ is the
Brewster angle for $H$-polarized waves and $\theta_C$ is the
critical angle.}
\end{figure}

\subsection{the isofrequency curve is a hyperbola with foci in the $k_z$ axis}
If the signs of the permittivity and permeability are of the form
$\boldsymbol{\varepsilon}=(-, +, +)$ and $\boldsymbol{\mu}=(+, -,
-)$, the isofrequency curve of the refracted wave vector is a
hyperbola with the foci in the $z$ axis, as shown in Fig.~6 (a).
\begin{figure}
\includegraphics[width=8cm]{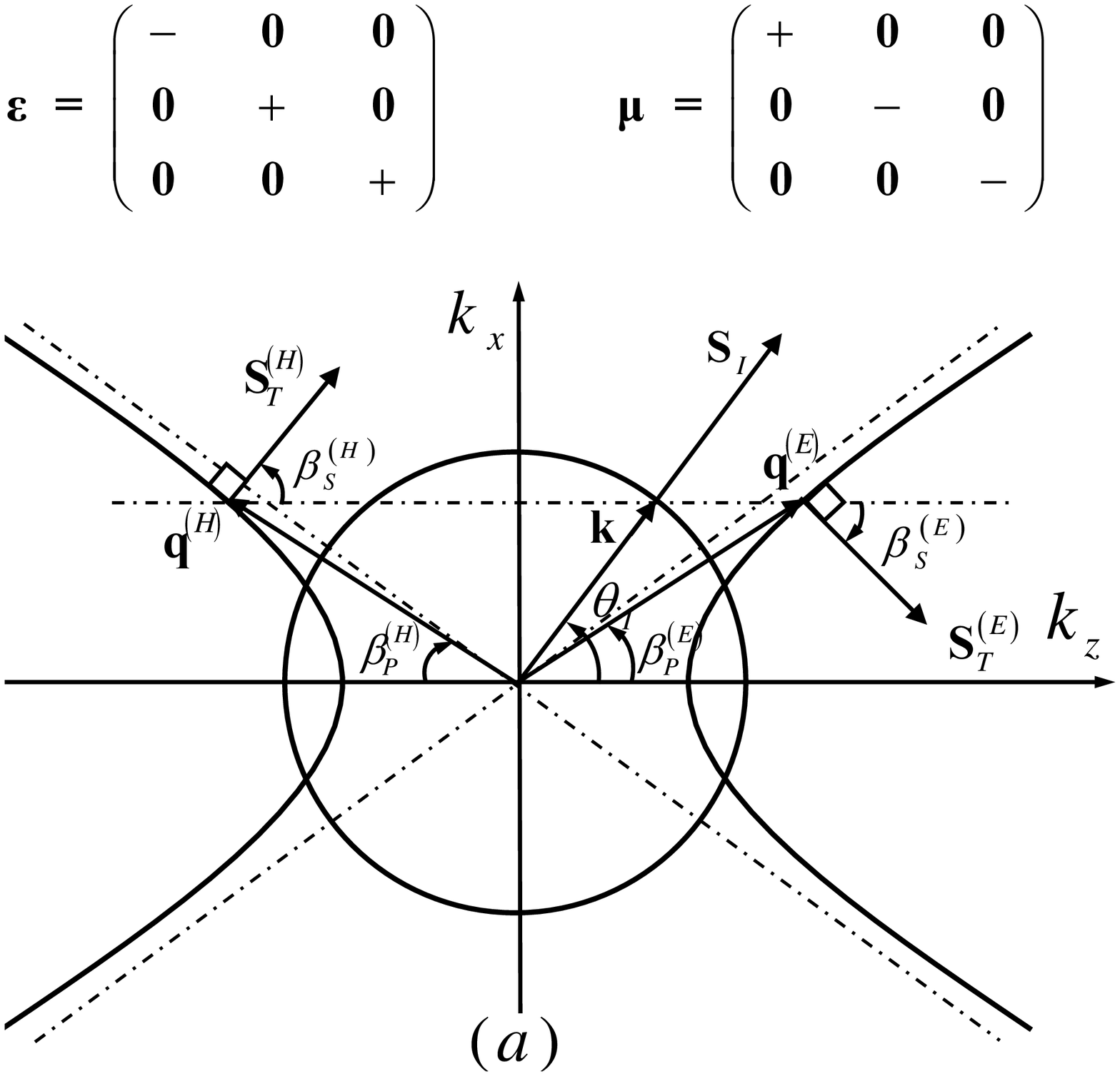}
\includegraphics[width=8cm]{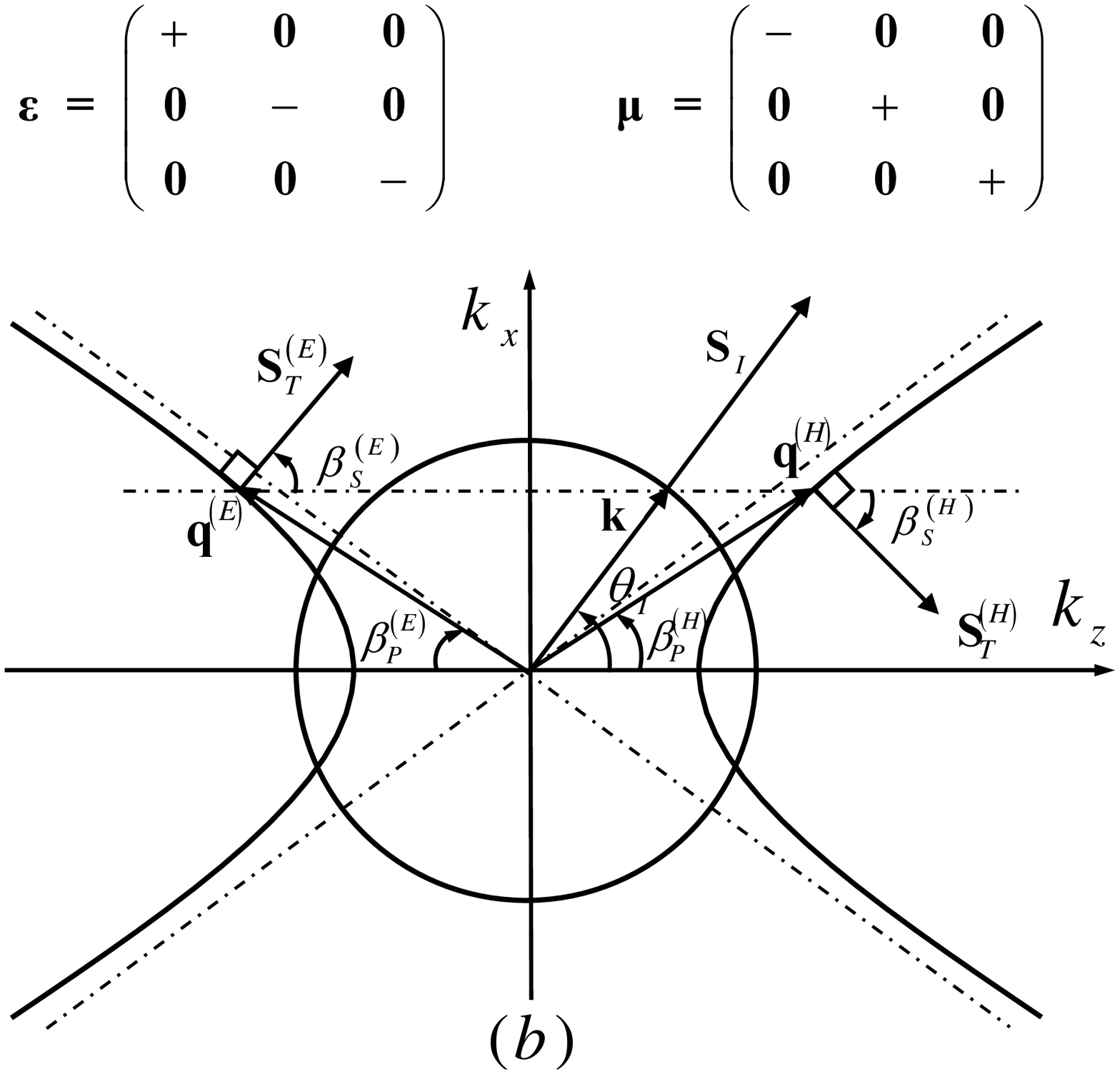}
\caption{\label{Fig5} The isofrequency curves to illustrate the
refraction. The circle and the hyperbola represent the  surfaces
of wave vectors in an isotropic regular material and the
anisotropic medium. In (a), $\beta_P^{(E)}=-\beta_P^{(H)}>0,
~\beta_S^{(E)}=-\beta_S^{(H)}<0$. In (b),
$\beta_P^{(H)}=-\beta_P^{(E)}>0, ~\beta_S^{(H)}=-\beta_S^{(E)}<0$.
There does not exist any critical angle. }
\end{figure}
Let us note that this case corresponds to the \textit{never
cutoff} indefinite medium in Ref.~\cite{Smith2003}. In view of the
discussions in the previous section, we have
\begin{eqnarray}
&&\mu_x>0,~q^{(E)}_z>0;~\mu_z<0,~{\bf q}^{(E)}_x\cdot{\bf S}_{T}^{(E)}<0.\\
&&\varepsilon_x<0,~q^{(H)}_z<0;~\varepsilon_z>0,~{\bf
q}^{(H)}_x\cdot{\bf S}_{T}^{(H)}>0.
\end{eqnarray}
For the $E$-polarized wave the wave vector is refracted regularly,
but the Poynting vector is refracted anomalously. For the
$H$-polarized wave the situation is just opposite. It is easy to
show that
\begin{equation}
\beta_P^{(E)}=-\beta_P^{(H)}>0, ~\beta_S^{(E)}=-\beta_S^{(H)}<0.
\end{equation}
Different from the case shown in Fig. 2 (a), there does not exist
any critical value for the incident angle, namely
\begin{equation}
0<|\theta_I|<\pi/2.
\end{equation}
As for the case of $\boldsymbol{\varepsilon}=(+, -, -)$ and
$\boldsymbol{\mu}=(-, +, +)$, the isofrequency curve of the
refracted wave vector is  shown in Fig. 6 (b). We can see that the
refraction behaviors of  $E$- and $H$-polarized waves are just
opposite to those in Fig. 6 (a). We give a numerical result of
$\boldsymbol{\varepsilon}=(-0.5, 0.4, 1)$ and
$\boldsymbol{\mu}=(1, -0.8, -2)$ in Fig.~7. We find that there
exists a Brewster angle for $E$-polarized waves.
\begin{figure}
\includegraphics[width=8cm]{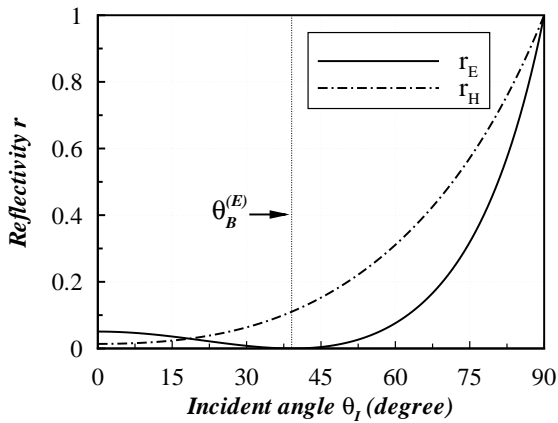}
\caption{\label{Fig1} Diagram of reflectivity as a function of
incident angle. Constitutive parameters of media are chosen as
$\varepsilon_I=1$, $\mu_I=1$, $\boldsymbol{\varepsilon}=(-0.5,
0.4, 1)$ and $\boldsymbol{\mu}=(1, -0.8, -2)$. The
$\theta_{B}^{(E)}$ is the Brewster angle for $E$-polarized waves.
}
\end{figure}

In the above discussions we find that $E$-polarized waves
propagate remarkably distinct from $H$-polarized waves in the
special anisotropic media. The two polarized waves are always
refracted into different directions. Even if the propagation
direction is along one principal axis, the two eigenmodes for one
propagation direction are always in two different directions,
\textit{i.e.}, $q_z^{(E)}=-q_z^{(H)}$ for one $q_x$. Then, the
refractive indices $n$ are always different for the two
polarizations and are relevant to the propagation directions.
Meanwhile, optic axes are the directions for which the two values
of the phase velocity $v_p=c/n$ are equal \cite{Chen,Born}.
Accordingly, there does not exist any optic axis in the
anisotropic medium, and the concept of ordinary waves or
extraordinary waves in uniaxial materials is not valid here. In
this sense, waves in the anisotropic medium all can be regarded as
extraordinary waves. The property that the two eigenmodes for one
propagation direction are always refracted into two different
directions indicates that the anisotropic media are
polarization-sensitive and can be put into potential applications,
such as dividing a beam into two pure polarizations. The
polarization beam splitter made of such media can divide beams
with larger splitting angles and splitting distances than
conventional ones \cite{Shiraishi1991}. In the above discussions
we have assumed that the principal axes are coincident with the
coordinate axes in which the media lie. If not so, new propagation
phenomena will occur except for those in the above and one can
investigate this issue by methods in
Refs.~\cite{Luo2005,Grzegorczyk2005a,Kong2000}.

Finally, we discuss how to realize the anisotropic media of a
unique dispersion. Several recent developments make the special
anisotropic media available. Firstly, metamaterials composed of
periodic arrays of split-ring resonators and conducting wires have
been demonstrated to be able to construct anisotropic media of the
unique dispersion relation
\cite{Smith2004a,Grzegorczyk2005b,Chen2005}. A more promising
choice is the metamaterial composed of periodic inductor-capacitor
loaded transmission line circuits because it has lower loss and
wider bandwidth
\cite{Eleftheriades2002,Iyer2006,Feng2005,Feng2006}. Such
metamaterials can also exhibit the above dispersion relations. In
addition, certain designs of photonic crystals have been shown to
be able to model the dispersion relation of anisotropic materials
\cite{Shvets2003,Shvets2004,Urzhumov2005}. Therefore, there is no
physical or technical obstacles to make the special anisotropic
media.

\section{conclusion}\label{sec4}
In summary, we have investigated the wave propagation in
anisotropic media for which the dispersion relation of any
polarized waves is the same. We have analysed the reflection and
refraction behaviors of electromagnetic waves at the interface
between an isotropic regular medium and the anisotropic media. We
show that in the anisotropic media, the refraction angles of both
phase and energy flow for $E$-polarized waves are opposite to the
counterparts for $H$-polarized waves, that is,
$\beta_P^{(E)}=-\beta_P^{(H)}$ and $\beta_S^{(E)}=
-\beta_S^{(H)}$, though the dispersion relation for $E$-polarized
waves is the same as that for $H$-polarized waves. The refraction
behaviors of wave vector and energy flow are significantly
different for a certain polarized wave. In addition, we find many
interesting characteristics of wave propagation in the medium.
Firstly, the wave propagation exhibits different behaviors with
different sign combinations of the permittivity and permeability
tensors. Secondly, the reflection coefficient becomes zero at
certain angles not only for H-polarized wave but also for
E-polarized wave. Therefore it is reasonable to extend the concept
of Brewster angle from H-polarized wave to E-polarized wave.
Lastly, the Brewster angle can be larger than the critical angle,
which is called the inversion of critical angle. Our results
indicate that it is necessary to study the propagation of the
phase and the energy for both $E$- and $H$-polarized waves in
order to obtain a complete knowledge on the wave propagation in
anisotropic media.

Due to the unique dispersion relation, characteristics of
electromagnetic wave propagation in the special anisotropic media
are remarkably distinct from those in isotropic or uniaxial LHM.
The most attracting one is that the anisotropic media are
polarization-sensitive, which can lead to many applications, such
as to fabricate polarization splitters with higher efficiencies
than conventional counterparts. Before we end this paper, we
stress that the special anisotropic media can be realized by
anisotropic metamaterials having been constructed in laboratory
\cite{Smith2004a,Grzegorczyk2005b,Feng2005,Feng2006}. Then the
properties of wave propagation in them can be studied
experimentally and favorable applications can be realized by the
anisotropic media.

\begin{acknowledgements}
We would like to thank Professor Yijun Feng for many helpful
discussions. This work was supported in part by the National
Natural Science Foundation of China (No.~10125521, 10535010) and
the 973 National Major State Basic Research and Development of
China (G2000077400).
\end{acknowledgements}


\begin{references}

\bibitem{Veselago1968}    V. G.  Veselago, Sov. Phys. Usp. {\bf 10},  509 (1968).
\bibitem{Pendry2000}   J. B. Pendry, \prl {\bf 85}, 3966 (2000).

\bibitem{Shelby2001} R. A. Shelby, D. R. Smith, and S. Schultz, Science {\bf 292},
77 (2001).

\bibitem{Parazzoli2003}  C. G. Parazzoli, R. B. Greegor, K. Li, B. E. C. Koltenbah,
 and M. Tanielian, \prl {\bf 90}, 107401 (2003).

\bibitem{Houck2003} A. A. Houck, J. B. Brock, and I. L. Chuang, \prl {\bf 90}, 137401
(2003).

\bibitem{Kong2002} J. A. Kong, B. Wu, and Y. Zhang, \apl {\bf 80}, 2084 (2002).

\bibitem{Smith2002} D. R. Smith, D. Schurig,  and J. B. Pendry, \apl {\bf 81}, 2713 (2002).

\bibitem{Markos2002} P. Markos and  C. M. Soukoulis, \pre {\bf 65} 036622 (2002).

\bibitem{Foteinopoulou2003} S. Foteinopoulou, E. N. Economous, and  C. M. Soukoulis, \prl {\bf 90}, 107402 (2003).

\bibitem{Notomi2000} M. Notomi, \prb {\bf 62}, 10696 (2000).

\bibitem{Luo2002a}  C. Luo, S. G. Johnson, J. D.  Joannopoulos, and J. B. Pendry, \prb {\bf 65}, 201104(R) (2002).

\bibitem{Luo2002b} C. Luo, S. G. Johnson, J. D. Joannopoulos, {\apl} {\bf 81}, 2352 (2002).

\bibitem{Luo2002c}  C. Luo, S. G. Johnson, J. D. Joannopoulos, and J. B. Pendry, {Optics Express}
{\bf 11}, 746 (2003).

\bibitem{Moussa} R. Moussa, S. Foteinopoulou, L. Zhang, G. Tuttle, K. Guven, E. Ozbay, and C. M.
Soukoulis, {\prb} {\bf 71}, 085106 (2005).

\bibitem{Zhang2003}   Y. Zhang, B. Fluegel, A. Mascarenhas, \prl {\bf 91}, 157404 (2003).

\bibitem{Lindell2001} I. V. Lindell, S. A. Tretyakov,
K. I. Nikoskinen, and S. Ilvonen, Microw. Opt. Technol. Lett. {\bf
31}, 129 (2001).

\bibitem{Luo2005} H. Luo, W. Hu, X. Yi, H. Liu, and J. Zhu, {\oc} {\bf 254},
353  (2005).

\bibitem{Hu2002}  L. B. Hu, S. T. Chui, \prb {\bf 66},
085108 (2002).

\bibitem{Zhou2003}  L. Zhou, C. T. Chan,  and P. Sheng, \prb {\bf 68}, 115424 (2003).

\bibitem{Shen2005} N. H. Shen, Q. Wang, J. Chen, Y. X. Fan, J. Ding, H. T. Wang, Y. Tian, and
N. B. Ming, \prb {\bf 72}, 153104 (2005).

\bibitem{Smith2004}  D. R. Smith, P. Kolinko, and D. Schurig, {J. Opt. Soc. Am. B} {\bf 21},
1032 (2004).

\bibitem{Grzegorczyk2005a}  T. M. Grzegorczyk, M. Nikku, X. Chen, B. Wu, and J. A. Kong,
IEEE Transactions on Microwave Theory and Techniques,  {\bf 53},
 1443 (2005).

\bibitem{Smith2003}  D. R. Smith and D. Schurig, \prl {\bf 90}, 077405 (2003).

\bibitem{Grzegorczyk2005b} T. M. Grzegorczyk, Z. M. Thomas, and J. A. Kong,  \apl {\bf 86}, 251909 (2005).

\bibitem{Smith2004a}  D. R. Smith, D. Schurig, J. J. Mock, and P. Kolinko, \apl {\bf 84}, 2244 (2004).

\bibitem{Eleftheriades2002}  G. V. Eleftheriades, A. K. Iyer, and P. C. Kremer, IEEE Trans.
Microwave Theory Tech. {\bf 50}, 2702 (2002).

\bibitem{Caloz2003} C. Caloz and T. Itoh, IEEE Microwave and Wireless Components Letters. {\bf 13},
547 (2003).

\bibitem{Caloz2004} C. Caloz and T. Itoh, IEEE Trans. Antennas Propag. {\bf 52}, 1159
(2004).

\bibitem{Grbic2003} A. Grbic and G. V. Eleftheriades, IEEE Antennas Propag. {\bf 51},
2604 (2003).

\bibitem{Iyer2006}  A. K. Iyer and G. V. Eleftheriades, {J. Opt. Soc. Am. B} {\bf 23},
553 (2004).

\bibitem{Cui2005}  T. J. Cui, Q. Cheng, Z. Z. Huang, and Y. Feng, {\prb} {\bf
72}, 035112 (2005).

\bibitem{Feng2005}  Y. Feng, X. Teng, Y. Chen, and T. Jiang, {\prb} {\bf 72}, 245107 (2005).

\bibitem{Chen} H. C. Chen, \textit{Theory of electromagnetic waves}, Chapter 5 (McGraw-Hill, New York, 1983).

\bibitem{Luo2006} H. Luo, W. Shu, F. Li, and Z. Ren, in review.

\bibitem{Kong2000} J. A. Kong, \textit{Electromagnetic wave theory} (EMW, New York, 2000).

\bibitem{Born} M. Born and E. Wolf, \textit{Principles of Optics}, 7th ed. (Cambridge, Cambridge, 1999).

\bibitem{Shiraishi1991} K. Shiraishi, T. Sato, S. Kawakami, \apl {\bf 58}, 211 (1991).

\bibitem{Chen2005} X. Chen, B. I. Wu, J. A. Kong, and T. M. Grzegorczyk,  \pre {\bf 71}, 046610 (2005).

\bibitem{Feng2006}  X. Teng, J. Zhao, T. Jiang, and Y. Feng, J. Phys. D {\bf 39}, 213
(2006).

\bibitem{Urzhumov2005} Y. A. Urzhumov and  G. Shvets, \pre {\bf 72}, 026608 (2005) .

\bibitem{Shvets2003} G. Shvets, \prb {\bf 67}, 035109 (2003).

\bibitem{Shvets2004}  G. Shvets, Y. A. Urzhumov, \prl   {\bf 93}, 243902 (2004).

\end{references}
\end{document}